\documentclass[%
 aip,
 amsmath,amssymb,
 reprint,
]{revtex4-1}

\usepackage{graphicx}
\usepackage{dcolumn}
\usepackage{bm}

\usepackage[utf8]{inputenc}
\usepackage[T1]{fontenc}
\usepackage{mathptmx}
\usepackage{etoolbox}

\makeatletter
\def\@email#1#2{
 \endgroup
 \patchcmd{\titleblock@produce}
  {\frontmatter@RRAPformat}
  {\frontmatter@RRAPformat{\produce@RRAP{*#1\href{mailto:#2}{#2}}}\frontmatter@RRAPformat}
  {}{}
}
\makeatother
\begin{document}

\preprint{AIP/123-QED}

\title{Complex network growth model: Possible isomorphism between nonextensive statistical mechanics and random geometry}

\author{Constantino Tsallis}%
\email{tsallis@cbpf.br}
\affiliation{Centro Brasileiro de Pesquisas F\'{\i}sicas and National Institute of Science and Technology for Complex Systems,  Rua Dr. Xavier Sigaud 150, 22290-180, Rio de Janeiro, Brazil}
\affiliation{Santa Fe Institute, 1399 Hyde Park Road, Santa Fe, 87501 NM, United States}
\affiliation{Complexity Science Hub Vienna, Josefst\"adter Strasse 39, 1080 Vienna, Austria}
 
\author{Rute Oliveira}%
\affiliation{Federal University of Rio Grande do Norte, Departamento de F\'isica Te\'orica e Experimental, Natal-RN, 59078-900, Brazil}

\date{\today}

\begin{abstract}
In the realm of Boltzmann-Gibbs statistical mechanics there are three well known isomorphic connections with random geometry, namely (i) the Kasteleyn-Fortuin theorem which connects the $\lambda \to 1$ limit of the $\lambda$-state Potts ferromagnet with bond percolation, (ii) the isomorphism which connects the $\lambda \to 0$ limit of the $\lambda$-state Potts ferromagnet with random resistor networks, and (iii) the de Gennes isomorphism which connects the $n \to 0$ limit of the $n$-vector ferromagnet with self-avoiding random walk in linear polymers. We provide here strong numerical evidence that a similar isomorphism appears to emerge connecting the energy $q$-exponential distribution $\propto e_q^{-\beta_q \varepsilon}$ (with $q=4/3$ and $\beta_q \omega_0 =10/3$) optimizing, under simple constraints, the nonadditive entropy $S_q$ with a specific geographic growth random model based on preferential attachment through exponentially-distributed weighted links,  $\omega_0$ being the characteristic weight. 
\end{abstract}

\maketitle

\begin{quotation}
Several examples exist of isomorphism between specific models in the realm of Boltzmann-Gibbs statistical mechanics with random geometry models. Such examples include the Kasteleyn-Fortuin theorem related with bond percolation,  the zero-state limit of the Potts ferromagnet related with random resistor networks, and the de Gennes isomorphism of the zero-component limit of the $\mathbf{n}$-vector model with self-avoiding random walk. We present here strong numerical evidence that the same happens in the realm of nonextensive statistical mechanics. Indeed, the energy distribution associated with a geographical $\mathbf{d}$-dimensional preferential-attachment-based (asymptotically) scale-free growth model is given by a simple $\mathbf{q}$-exponential with $\mathbf{q=4/3}$.
\end{quotation}

\section{Introduction} \label{introduction}
Within Boltzmann-Gibbs (BG) statistical mechanics, the energy distribution of a Hamiltonian system in thermal equilibrium at temperature $T$ is given by the following matrix density $\rho$ \cite{Huang1987,Balian1991}:
\begin{equation}
\rho({\cal H})=\frac{e^{-\beta {\cal H}}}{Z(\beta)} \,,
\end{equation}
where $\beta \equiv 1/kT$ and the partition function is defined as $Z(\beta) \equiv Tr \,e^{-\beta {\cal H}}$, ${\cal H}$ being the Hamiltonian. In their diagonalized version, these quantities become
\begin{equation}
p(E_i)=\frac{e^{-\beta E_i}}{Z(\beta)} \,,
\end{equation}
where $Z(\beta) \equiv \sum_i e^{-\beta E_i}$, $E_i$ being the $i$-th eigenvalue of the Hamiltonian ${\cal H}$. 

Occasionally, there are models whose BG thermostatistical approach is isomorphic to random geometrical models. Let us mention three well known such examples: (i) The  Kasteleyn-Fortuin theorem \cite{KasteleynFortuin1969} which connects bond percolation in an arbitrary graph or lattice with the $\lambda \to 1$ limit of the $\lambda$-state Potts ferromagnetic model in the same graph (see \cite{TsallisMagalhaes1996} and references therein); (ii) The random resistor (or impedance) network which connects the Ohmic behavior of an arbitrary resistor graph with the $\lambda \to 0$ limit of the just mentioned $\lambda$-state Potts ferromagnetic model in the same graph (see \cite{TsallisMagalhaes1996,TsallisConiglioRedner1983} and references therein); (iii) The de Gennes isomorphism \cite{deGennes1972,MadrasSlade2013} which connects self-avoiding random walk on an arbitrary graph (equivalently the growth configurations of a linear polymer) with the $n \to 0$ limit of the $n$-vector ferromagnet on the same graph. We briefly review these three isomorphisms in Section \ref{BGmodels}. 

In the present paper (Section \ref{network}), we describe and numerically study a random geometrical model, namely the growth of an asymptotically scale-free geographic weighted-link preferential-attachment network. Its numerical study provides a strong indication of being isomorphic to a simple thermostatistical model within nonextensive statistical mechanics \cite{Tsallis1988,Tsallis2009,Tsallis2019}. 

\section{Isomorphic models within the BG statistical mechanics} \label{BGmodels}
We briefly review here three well known examples of isomorphism between models within BG thermostatistics and nontrivial random geometrical models. We refer to the $\lambda\to1$ and $\lambda\to 0$ limits of the $\lambda$-state Potts ferromagnet, and the $n\to 0$ limit of the $n$-vector ferromagnet.
\subsection{The Potts ferromagnet}
The Hamiltonian of the $\lambda$-state Potts ferromagnetic model \cite{Potts1951} is defined as follows:
\begin{equation}
{\cal H} =-\lambda \sum_{i,j} J_{ij}\delta_{\sigma_i,\sigma_j} \;\;\;(\sigma_i=1,2,\dots,\lambda,\,\forall i)\,,
\end{equation}
where $J_{i,j} >0, \forall (i,j)$, and the sum runs over all pairs of ``spins" located at the sites of an arbitrary lattice (finite or infinite, regular or not, translationally invariant, i.e., Bravais lattice, scale invariant, i.e., fractal lattice, etc.) and $\delta_{\sigma_i,\sigma_j}$ is the Kronecker's delta. The particular case $\lambda = 2$ is (through a trivial energy shift in the Hamiltonian) identical to the standard spin $1/2$ Ising model.  The elementary Potts interaction (single bond) yields a two-level spectrum: one level with energy $-\lambda J_{ij}$ and degeneracy $\lambda$, and the other one with energy $0$ and degeneracy $\lambda (\lambda -1)$. To every single bond between sites $i$ and $j$, we may associate its {\it thermal transmissivity} (see \cite{TsallisMagalhaes1996} and references therein)
\begin{equation}
t_{ij} \equiv \frac{1-e^{-\lambda J_{ij}/kT}}{1+(\lambda-1)e^{-\lambda J_{ij}/kT}} \in [0,1]\,.
\label{tdefinition}
\end{equation}
Let us first consider a {\it series} array of two bonds or links (with Potts coupling constants $J_1$ and $J_2$) and three
vertices or sites. The transmissivity after tracing over the $\lambda$ states of the internal site is given by
\begin{equation}
t_s=t_1t_2 \,\;\;\;(series)\,.
\label{tseries}
\end{equation}
If we have instead a {\it parallel} array of two bonds (again with Potts coupling constants $J_1$ and $J_2$), the resulting coupling constant is given by $J_1+J_2$, which straightforwardly leads to
\begin{equation}
t_p=\frac{t_1+t_2+(\lambda-2)t_1t_2}{1+(\lambda-1)t_1t_2} \,\;\;\;(parallel)
\label{tparallel}
\end{equation}
(or, equivalently, $\frac{1-t_p}{1+(\lambda-1)t_p}=\frac{1-t_1}{1+(\lambda-1)t_1} \frac{1-t_2}{1+(\lambda-1)t_2}$). 
If we have two-open-site arrays which are not reducible to sequences of series and parallel operations (e.g., the Wheatstone bridge), we may use the Break-Collapse Method (see \cite{TsallisMagalhaes1996} and references therein) to calculate the equivalent transmissivity. 
\subsubsection{The $\lambda\to 1$ limit}
If we consider the analytic $\lambda\to 1$ limit Eq. (\ref{tseries}) remains as it stands, whereas Eq. (\ref{tparallel}) becomes
\begin{equation}
t_p=t_1+t_2-t_1t_2 \,\;\;\;(parallel)
\label{percolationparallel}
\end{equation}
(or, equivalently, $(1-t_p)=(1-t_1)(1-t_2)$). 
We notice that Eqs. (\ref{tseries}) and (\ref{percolationparallel}) are precisely the composition laws of independent probabilities. This is the basis of the Kasteleyn-Fortuin theorem \cite{KasteleynFortuin1969}, which rigorously establishes the isomorphism of the $\lambda\to 1$ Potts ferromagnet in an arbitrary lattice with bond percolation in the same lattice.

\subsubsection{The $\lambda\to 0$ limit}
If we define now 
\begin{equation}
t_i\equiv 1-\frac{g_0}{g_i} \,,
\end{equation}
where $g_0$ is some reference electrical conductance (i.e., the inverse of a reference electrical resistance), we straightforwardly verify that, in the $g_0/g_i \to 0$ limit, Eqs. (\ref{tseries}) and (\ref{tparallel}) become respectively (see, for instance, \cite{TsallisConiglioRedner1983})
\begin{equation}
g_s=\frac{g_1g_2}{g_1+g_2}\;\;\;(series)
\end{equation}
(or, equivalently $\frac{1}{g_s}=\frac{1}{g_1}+\frac{1}{g_2}$)
and
\begin{equation}
g_p=g_1+g_2\;\;\;(parallel)\,.
\end{equation}
These composition laws are precisely those of Ohmic conductances. If we have two-open-site arrays which are not reducible to sequences of series and parallel operations, we may use once again the Break-Collapse Method \cite{TsallisConiglioRedner1983} to calculate the equivalent conductance. 
This constitutes the basis of the isomorphism of the $\lambda\to 0$ Potts ferromagnet in an arbitrary lattice with random resistors in the same lattice.

\subsection{The $n$-vector ferromagnet}
The Hamiltonian of the $n$-vector or $O(n)$ ferromagnetic model  can be defined as follows \cite{Stanley1968}:
\begin{eqnarray}
{\cal H} &=&-\sum_{ij}J_{ij} \vec s_i .  \vec s_j \nonumber \\
&=&-\sum_{ij}J_{ij} \sum_{k=1}^n s_i^{(k)} s_j^{(k)}  \;\Bigl(\,\sum_{k=1}^n [s_i^{(k)}]^2=1 \,, \forall i \Bigr)
\end{eqnarray}
where $J_{i,j} >0, \forall (i,j)$, and the first sum runs over all pairs of spin located at the sites of an arbitrary lattice (finite or infinite, regular or not, translationally invariant, i.e., Bravais lattice, scale invariant, i.e., fractal lattice, etc.). The particular case $n = 1$ corresponds to the Ising model; $n=2$ corresponds to the XY model; $n=3$ corresponds to the Heisenberg model; $n\to\infty$ corresponds to the spherical model. In 1972 de Gennes proved \cite{deGennes1972} that the analytical extension $n\to 0$ is isomorphic to the growth of a self-avoiding linear polymer in the same lattice. This isomorphism certainly constitutes one of the landmarks of polymer physics.

\section{Possible isomorphic model within nonextensive statistical mechanics} \label{network}
\subsection{The random geometric weighted-link preferential-attachment growth model}
The growing $d$-dimensional network we focus on here has been introduced and studied in \cite{OliveiraBritoSilvaTsallis2021}, which we follow now. We start with one site at the origin. We then stochastically locate a second site (and then a third, a fourth, and so on up to N) through a probability  $p(r) \propto 1/r^{d+ \alpha_G} \;\;\; (\alpha_G > 0)$, 
where $r\geq1$ is the Euclidean distance from the newly arrived site to the center of mass of the pre-existing cluster; $\alpha_G$ is the {\it growth} parameter and $d=1,2,3$ is the dimensionality of the system (large $\alpha_G$ yields geographically concentrated networks). 

The site $i=1$ is then linked to the site $j=2$. We sample a random number $w_{ij}$ from a distribution $P(w)$ that will give us the corresponding link weight. Each site will have a \emph{total energy} $\varepsilon_i$ that will depend on how many links it has, noted $k_i$, and the widths $\{ w_{ij}\}$ of those links. At each time step, the site $i$ only has access to its \emph{local energy} $\varepsilon_i$ defined as:

\begin{eqnarray}\label{eq:energy_site}
\varepsilon_i \equiv \sum_{j=1}^{k_i} \frac{w_{ij}}{2} \;\;\;(w_{ij} \ge0)
\end{eqnarray}
The value of $\varepsilon_i$ will directly affect the probability of the site $i$ to acquire new links. Indeed, from this step on, the sites $i =3,4, ...$ will be linked to the previous ones with probability
\begin{eqnarray}
\Pi_{ij}\propto \frac{\varepsilon_{i}}{d^{\,\alpha_A}_{ij}} \;\;(\alpha_A \ge 0)\,,
\label{attachment}
\end{eqnarray}
where $d_{ij}$ is the Euclidean distance between $i$ and $j$, where $j$ runs over all sites linked to the site $i$. The {\it attachment} parameter $\alpha_A$ controls the importance of the distance in the preferential attachment rule (\ref{attachment}). When $\alpha_A \gg 1$ the sites tends to connect to close neighbours, whereas $\alpha_A \simeq 0$ tends to generate distant connections all over the network. Notice that, while the network size increases up to $N$ nodes, the variables $k_i$ and $\varepsilon_i$ (number of links and \emph{total energy} of the $i$-th node; $i=1,2,3 \dots,N$) also increase in time.

If we consider the particular case $P(w)=\delta(w-1)$, where $\delta(z)$ denotes the Dirac delta distribution, Eq. (\ref{attachment}) becomes $\Pi_{ij}\propto k_{i}/d^{\,\alpha_A}_{ij} \;\;(\alpha_A \ge 0)\,$, 
thus recovering the usual preferential attachment rule (see, for instance, \cite{BritoSilvaTsallis2016,CinardiRapisardaTsallis2020} and references therein). Note that, if we additionally consider the particular case $\alpha_A=0$, we recover the standard Barabasi-Albert model with $\Pi_i \propto k_i$~\cite{BarabasiAlbert1999, AlbertBarabasi2002}.

We are considering here the case where $w$ is given by the following stretched-exponential distribution:

\begin{eqnarray}\label{eq:w}
P(w) = \frac{\eta}{w_0\,\Gamma\left(\frac{1}{\eta}\right)} e^{-(w/w_0)^{\eta}} \;\;(w \ge 0; \,w_0 >0;\, \eta > 0)\,,
\end{eqnarray}
which satisfies $\int_0^\infty dw\,P(w)=1$.
As particular cases of Eq.~(\ref{eq:w}) we have: $\eta = 1$, which corresponds to an exponential distribution, $\eta = 2$, which corresponds to a half-Gaussian distribution, and $\eta \to \infty $, which corresponds to an uniform distribution within $w \in [0,w_0] $. Our aim here is to specifically study the $\eta=1$ case, i.e., $p(w)=w_0^{-1}e^{-w/w_0}\;(w \ge 0;\, w_0>0)$. In Fig. \ref{fig1} we present numerical results for $\alpha_A=d=2$ and increasing values of $N$. These results remain in fact the same for any $(\alpha_A,d)$ values such that $0 \le \alpha_A/d \le 1$. This is illustrated in Fig. \ref{fig2} for $\alpha_A/d=1\;(d=1,2)$.

\subsection{Finite-size effects}
\begin{figure}
\hspace{-0.9cm}
\includegraphics[width=8.0cm]{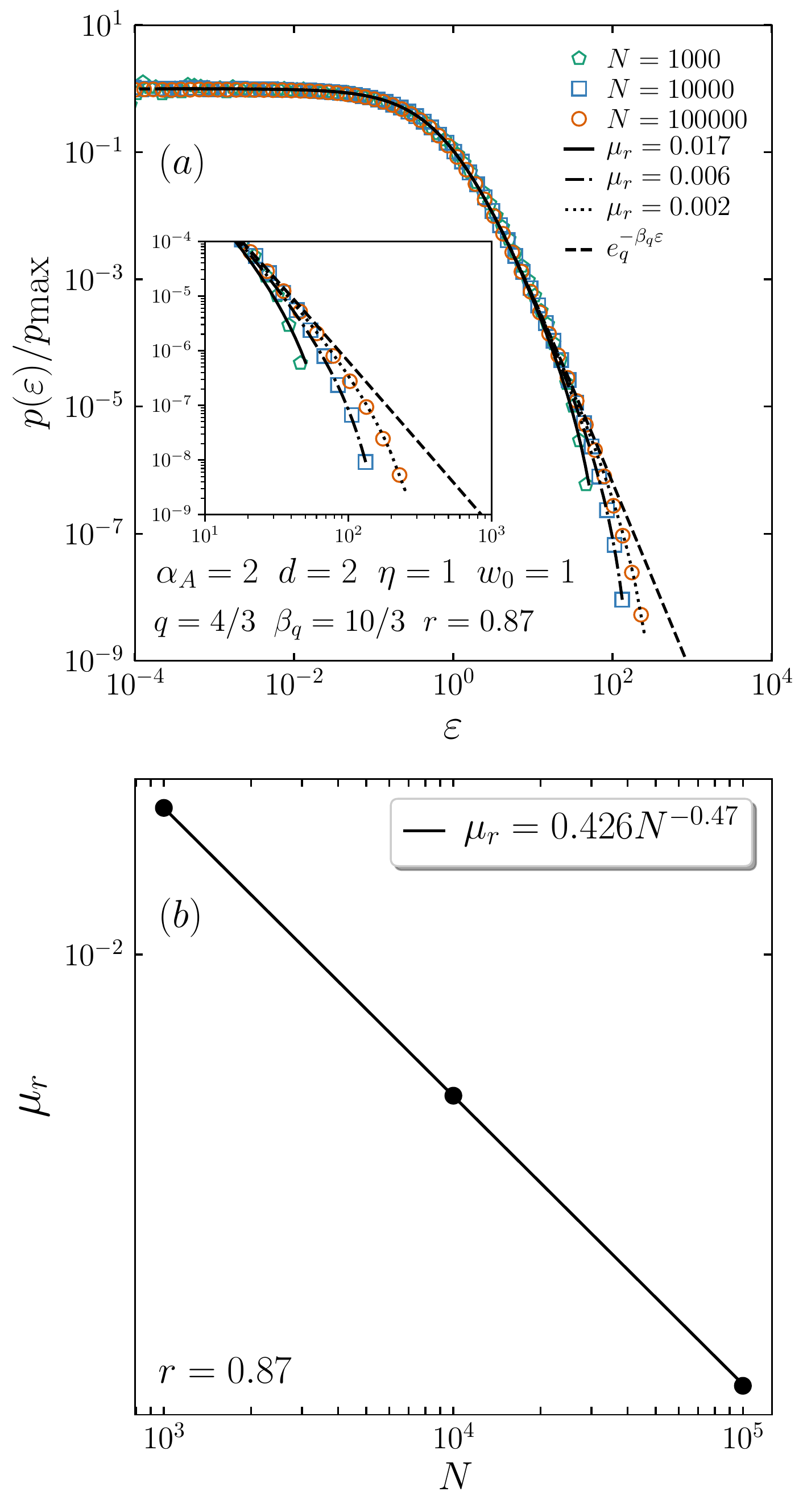}
\vspace{-0.2cm}
\caption{(a) The $\varepsilon$-dependence of the normalized probability $\xi(\varepsilon) \equiv p(\varepsilon)/p(0)$ corresponding to $(\eta,w_0,\alpha_A,d)=(1,1,2,2)$ and three typical values of $N$. The curves joining the points indicated for each value of $N$ are produced using $\xi(\varepsilon)$ given by Eq. (\ref{hyper}) with $r=0.87$ and the corresponding values of $\mu_r$. The dashed curve corresponds to the conjectured $N\to\infty$ limit $\xi(\varepsilon)=e_q^{-\beta_q \varepsilon}$ with $(q, \beta_q)=(4/3,10/3)$. (b) The $N$-dependence of $\mu_r$, where the fitting parameter $r$ has been chosen so that a straight line emerges in log-log representation, precisely corresponding here to $\mu_r=0.426N^{-0.47}$.
}
\label{fig1}
\end{figure}

\begin{figure}
\hspace{-0.2cm}
\includegraphics[width=8.0cm]{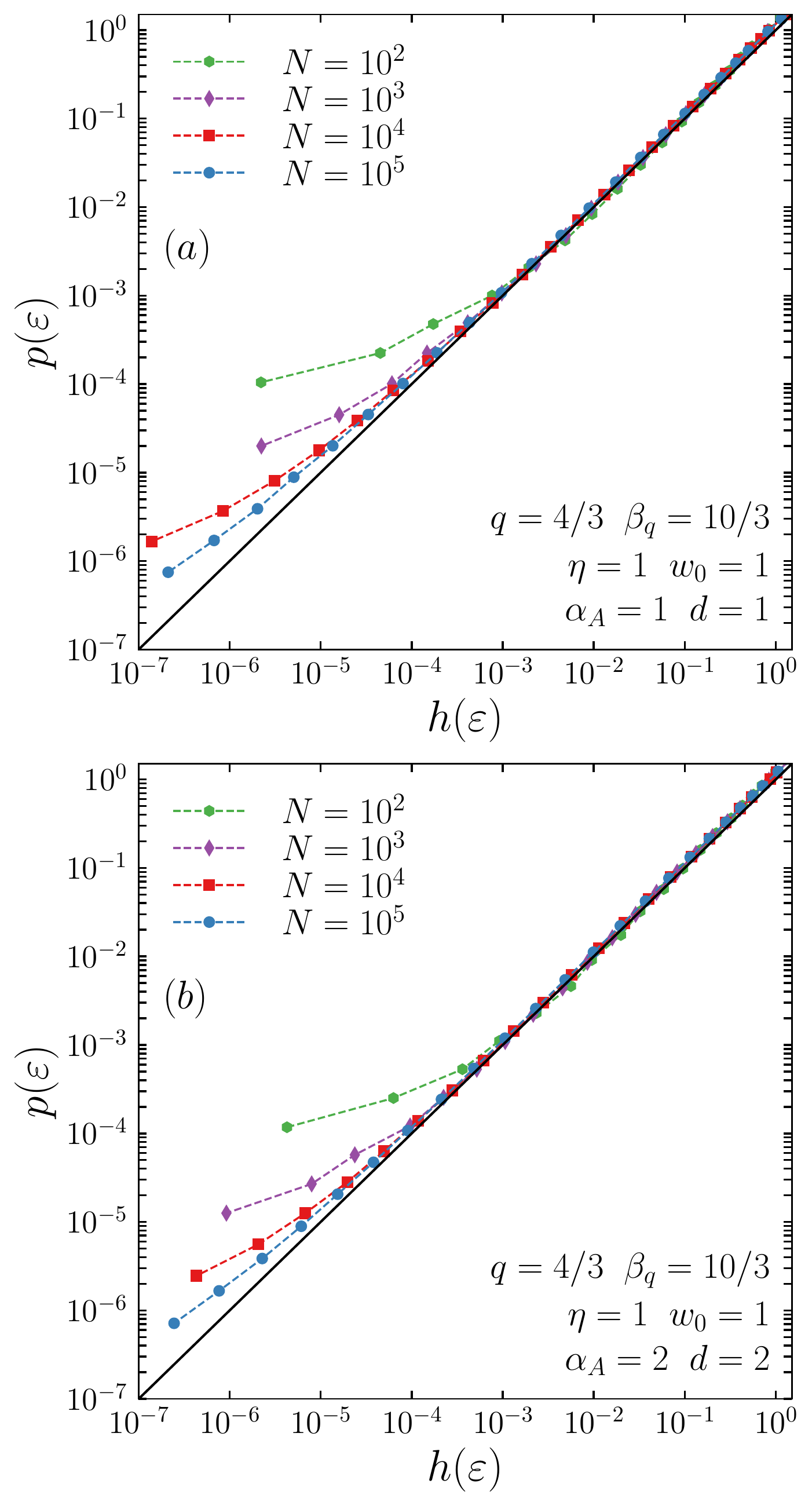}
\caption{The probability $p(\varepsilon)$ as a function of the corresponding histogram $h(\varepsilon)$ for $(\eta, w_0)=(1,1)$ and typical values of $N$ with  $\alpha_A=d=1$ (a) and $\alpha_A=d=2$ (b). Comparison of the histogram $h(\varepsilon)$ of site energies  with the nearly $q$-exponential distribution $p(\varepsilon)$. The dashed lines are guides to the eye. The bisector straight line corresponds to the conjectured $N \to\infty$ limit, namely $p(\varepsilon)=p(0)\,e_q^{-\beta_q  \varepsilon}$ with $(q,\beta_q)=(4/3,10/3)$.
}
\label{fig2}
\end{figure}

As we verify in both Figs. \ref{fig1} and \ref{fig2}, there are sensible effects on $p(\varepsilon)$ coming from the finiteness of $N$, which seemingly disappear in the $N\to\infty$ limit.
Following along the lines of \cite{TsallisBemskiMendes1999}, we check that these finite-size effects are satisfactorily described by the following equation:  
\begin{equation}
\frac{d\xi}{d\varepsilon}=-\mu_r \xi^r -(\beta_q-\mu_r)\xi^q \;\;(r \le q;\varepsilon \ge 0) \,,
\end{equation}
where $\xi(\varepsilon)\equiv p(\varepsilon)/p^{max} \in [0,1]$. Consequently
\begin{eqnarray}
\varepsilon&=&\int_\xi^1 \frac{dx}{\mu_r x^r+(\beta_q - \mu_r) x^q} \nonumber \\
&=&\frac{1}{\mu_r}\int_\xi^1dx \Bigl\{ \frac{1}{x^r}-   \frac{(\beta_q/\mu_r-1)x^{q-2r)}}{1+(\beta_q/\mu_r -1)x^{q-r}}  \Bigr\}    \nonumber \\
&=&\frac{1}{\mu_r}  \Bigl\{ \frac{\xi^{1-r}-1}{r-1}-\frac{(\beta_q/\mu_r)-1}{1+q-2r}  \nonumber \\
&&\times \Bigl[ H(1;q-2r,q-r,(\beta_q/\mu_r)-1)  \nonumber \\
&&- H (\xi;q-2r,q-r,(\beta_q/\mu_r)-1)     \Bigr]  
\Bigr\} 
\label{hyper}
\end{eqnarray}
with
\begin{equation}
H(\xi; a,b,c) \equiv \xi^{1+a} F\Bigl( \frac{1+a}{b},1;\frac{1+a+b}{c};-\xi^b c \Bigr)
\end{equation}
where $F$ is the hypergeometric function. We can verify for instance that, for $1<r<q$ and $0<\mu_r \ll \beta_q$, three regions emerge. These three regions are characterized as follows: 
for small $\varepsilon$, $\xi$ is nearly constant; for intermediate $\varepsilon$, $\xi$ decreases as $1/\varepsilon^{1/(q-1)}$; finally, for large $\varepsilon$, $\xi$ further decreases, now as $1/\varepsilon^{1/(r-1)}$. If $r\le 1$, this function vanishes even faster for increasing $\varepsilon$ (see details in \cite{TsallisBemskiMendes1999}).


\section{Conclusion}
The present numerical results provide a strong indication that, in the $N\to\infty$ limit, $\mu_r$ vanishes. Consequently, the differential equation $\frac{d\xi}{d\varepsilon}=- \beta_q\xi^q \;\;(\varepsilon \ge 0)$ is expected to be satisfied, hence 
\begin{equation}
p(\varepsilon)=p(0)\,e_q^{-\beta_q \, \varepsilon}\,.
\end{equation} 
Let us emphasize at this point that this distribution precisely is the one which optimizes, under simple constraints, the nonadditive entropy $S_q =k\frac{1-\sum_i p_i^q}{q-1}\;(S_1=S_{BG}=-k\sum_i p_i\ln p_i;\, \sum_ip_i=1)$ \cite{Tsallis1988,Tsallis2009,Tsallis2019}.

From the discussion presented in Fig. \ref{fig2}, as well as a variety of numerical checks that we have concomitantly performed, we are allowed to conjecture that $(q,\beta_q)=(4/3,10/(3w_0))$ for $\alpha_A/d \in [0,1]$. Therefore, the $N\to\infty$ limit of the random geometry growth model that we have focused on here appears to be isomorphic to a simple model within nonextensive statistical mechanics, more precisely the model whose total energy is just the sum of all the site energies $\{\varepsilon_i\}$.  This connection obviously is fully analogous to those three described in Section \ref{BGmodels} within Boltzmann-Gibbs statistical mechanics. Its rigorous proof remains to be done.\\

\begin{acknowledgments}
We acknowledge useful remarks from L.R. da Silva and S. Brito. Also, we have benefited from partial financial support by the Brazilian agencies CNPq, Capes and Faperj. We also thank the High Performance Computing Center (NPAD/UFRN) for providing computational 
resources.
\end{acknowledgments}

\section*{Data Availability Statement}
The data that support the findings of this study are available from the corresponding author upon reasonable request.

\end{document}